\documentclass[12pt]{article}
\usepackage{graphicx}
\usepackage{float}
\begin{document}

\title{\bf Ricci Dark Energy of Amended FRW Universe in Chern-Simon Modified Gravity}

\author{M. Jamil Amir\thanks{mjamil.dgk@gmail.com}$~~$and Sarfraz Ali\thanks{sarfraz270@yahoo.com}
\\Department of Mathematics, University of Sargodha,\\Pakistan,\\
Department of Mathematics, University of Education, Lahore,\\Pakistan.\\}

\date{}

\maketitle

\begin{abstract}
The energy density of the universe is proportional to the Ricci scalar curvature in the dynamical
Chern-Simon (CS) modified gravity. In this paper, we consider the Amended Friedman-Robertson-Walker (AFRW)
universe and explore its scale factor and the Ricci Dark Energy. THese turned out to be well-defined and
definite. We compare the scale factors of FRW \cite{[17]}, Generalized Chaplygin gas (GCG) \cite{[18]}
and AFRW models graphically. The combined graph of these models show that the behavior of both FRW and
AFRW models is similar as these overlap each other for choosing particular values of the integration
constants. Also, we draw a combined graph of the Ricci dark energy densities of FRW and AFRW models, in
CS gravity, and the energy density of GCG. It shows that the densities of former two models are increasing
with time while the energy density of GCG is decreasing.

\end{abstract}


\section{Introduction}

The observed current accelerated expansion of the universe \cite{[1]}-\cite{[3]} is an interesting and
perplexing problem that has become a serious challenge for gravity theories. The astrophysical observations
have provided the astonishing result that around $95-96$ percent of the content of the universe is in
the form of dark matter (DE) plus Dark Energy (DE), with only about $4-5$ percent being represented by byronic
matter \cite{[4]}. More intriguingly, around $70$ percent of the energy-density is in the form of
what is called "Dark Energy", and is responsible for the acceleration of the distant type Ia supernovae \cite{[5]}.
Hence, today's models of astrophysics and cosmology face two fundamental problems, the DE problem
and the dark matter problem, respectively. Although, in recent years different
suggestions have been proposed to overcome these issues, a satisfactory answer has yet to be obtained.
A very promising way to explain the observational data is to assume that at large scales the
Einstein gravity model of general relativity (GR) breaks down and a more general action describes
the gravitational field.

Theoretical models in which the standard Einstein-Hilbert action was modified by
replacing the Ricci scalar R with f(R), an arbitrary function of the Ricci scalar R, pioneerily proposed in \cite{[6]}.
Consequently, a considerable efforts to modify gravity have been made in order to accommodate this large
distance observation. A promising extension of GR is Chern-Simons (CS) modified gravity
\cite{[7]}-\cite{[9]} in which the Einstein-Hilbert action is modified by adding a parity-violating CS
term which couples to gravity using scalar field. It is interesting to note that the CS
correction introduces parity violation through a pure curvature term, as is usually considered
in GR. In fact, CS modified gravity can be obtained explicitly from superstring
theory, where the CS term in the Lagrangian density is essential due to the Green-Schwarz anomaly-canceling
mechanism, upon four-dimensional compactification  \cite{[10]}.

Two formulations of CS modified gravity exist as independent theories, namely, the nondynamical formulation and
the dynamical formulation  \cite{[9]}. In the former case, the CS scalar is a prior prescribed function, where its
effective evolution equation reduces to a differential constraint on the space of allowed solutions; in the latter case,
the CS is treated as a dynamical field, possessing an effective stress-energy tensor and an evolution equation.

The action of dynamical CS modified gravity consists of the Einstein-Hilbert term plus the product
of a scalar field and the Pontryagin density (the contraction of the Riemann curvature tensor with its dual),
plus the action for this scalar field or other matter fields or both. The correction proportional to the Pontryagin
density modifies the field equations for the metric components by adding two extra terms to the Einstein field equations,
called C-tensor (C-tensor) and the stress-energy tensor for scalar field. The C-tensor depends on derivatives
of the CS scalar and the contraction of the Levi-Civita tensor with covariant derivatives of the Ricci tensor and
the dual Riemann tensor. In addition, the variation of the action with respect to the CS scalar field leads to an
equation of motion for this field, which is sourced by the Pontryagin density.

In literature, a number of DE models are available which have been built
on the holographic principle, known as holographic DE models \cite{[11]}, \cite{[12]}. For these models,
the energy density of a given system is proportional to the inverse square of characteristic
length of the system \cite{[13]}. Although, these DE models exhibit the problem of
fine tuning or coincidence problem. Chattopadhyay et al. \cite{[21]} considered the modified and the extended
holographic Ricci DE models in fractal universe. They proved that the modified holographic dark energy density
is decreasing with cosmic time and extended holographic dark energy density is increasing with time.

Gao et al. \cite{[14]} proposed a DE model
whose characteristic length is the average radius of the Ricci scalar, $R^{\frac{-1}{2}}$. Thus, the Ricci DE density
is proportional to the Ricci scalar, i.e.,
\begin{eqnarray}
\rho_{x}=-\frac{\alpha}{16\pi}R,
\end{eqnarray}
where R represent Ricci scalar and $\alpha$ is a constant to be determined.

This paper is devoted to explore the Ricci DE of AFRW universe in the frame work of CS modified gravity. Also,
the graphical comparison of our results with those of FRW and GCG models will be presented.

\section{Dark Energy Model}
The line element of AFRW universe is given by  \cite{[15]}
\begin{eqnarray}
ds^{2}=-a^{2}(t)c^{2}dt^{2}+a^{2}(t)[\frac{1}{1-k r^{2}}dr^{2}+r^{2}(d\theta^{2}+\sin^{2}{\theta}d\phi^{2})].
\end{eqnarray}
The Ricci scalar for this metric is evaluated as
\begin{eqnarray}
R=-6(\frac{\ddot{a}}{a^{3}c^{2}}+\frac{k}{a^{2}}),
\end{eqnarray}
where $a(t)$ represents scale factor and double dot is used for second order derivative w.r.t. $"t"$.
Making use of Eq.(3) in Eq.(1), the Ricci DE density turns out to be
\begin{eqnarray}
\rho_{x}=\frac{3\alpha}{8\pi}(\frac{\ddot{a}}{a^{3}c^{2}}+\frac{k}{a^{2}}).
\end{eqnarray}
Now, we discuss this Ricci DE model in the framework of dynamical CS modified gravity.
The action describing CS modified gravity theory is, given as
\begin{eqnarray}
S=\frac{1}{16\pi G}\int d^{4}x[\sqrt{-g}R+\frac{l}{64\pi}\Theta~^{*}RR-\frac{1}{2}\partial^{\mu}\Theta\partial_{\mu}\Theta)]+S_{mat},
\end{eqnarray}
where $^{*}RR$ is called Pontryagin term which is topological invariant, $l$ is coupling constant,
$\Theta$ is dynamical variable, $S_{mat}$ is the action of matter. The variation of action, given in Eq.(5), with respect to
metric $g_{\mu\nu}$ and the scalar field $\Theta$ give the following respective field equations in CS modified gravity
\begin{eqnarray}
G_{\mu\nu}+lC_{\mu\nu} &=&8\pi G T_{\mu\nu},
\\
g^{\mu\nu}\nabla_{\mu}\nabla_{\nu}\Theta &=&-\frac{l}{64\pi} ~^{*}RR,
\end{eqnarray}
where $G_{\mu\nu}$ is the Einstein tensor and $C_{\mu\nu}$ is the C-tensor.
The vanishing of the C-tensor depends on the type of relation between the Ricci tensor and the
energy-momentum tensor of matter. A 3-dimensional space is conformally flat if the Cotton
tensor vanishes. If matter is present, the Ricci tensor is related to the energy-momentum tensor
of matter by means of the Einstein field equations. Then the vanishing of the C-tensor imposes severe
restrictions on the energy-momentum tensor. The C-tensor also plays an important role in the context of the
Hamiltonian formulation of GR.
The C-tensor can be written as
\begin{eqnarray}
C^{\mu\nu}=-\frac{1}{2\sqrt{-g}}[\upsilon_{\sigma}\epsilon^{\sigma\mu\alpha\beta}\nabla_{\alpha}R^{\nu}_{\beta}+
\frac{1}{2}\upsilon_{\sigma\tau}\epsilon^{\sigma\nu\alpha\beta}R^{\tau\mu}_{\alpha\beta}]+(\mu\longleftrightarrow\nu),
\end{eqnarray}
where
\begin{eqnarray}
\upsilon_{\sigma}\equiv\nabla_{\sigma}\Theta,~~~~~ \upsilon_{\sigma\tau}\equiv\nabla_{\sigma}\nabla_{\tau}\Theta.
\end{eqnarray}
Again from Eq.(7), the Pontryagin term is expressed as
\begin{eqnarray}
^{*}RR=~^{*}R^{a}~_{b}~ ^{cd}~R^{b}~_{acd},
\end{eqnarray}
where $R^{b}~_{acd}$ is the Riemann tensor and $^{*}R^{a}~_{b}~ ^{cd}$ is the dual Riemann tensor defined as
\begin{eqnarray}
^{*}R^{a}~_{b}~ ^{cd}=\frac{1}{2}\epsilon^{cdef}R^{a}~_{bef}.
\end{eqnarray}
The energy-momentum tensor $T_{\mu\nu}$ is divided into two parts such that
\begin{eqnarray}
T_{\mu\nu}=T^{RDE}_{\mu\nu}+T^{\Theta}_{\mu\nu},
\end{eqnarray}
where $T^{\Theta}_{\mu\nu}$ is the energy-momentum tensor associated with the scalar field $\Theta$ defined as
\begin{eqnarray}
T^{\Theta}_{\mu\nu}=(\nabla_{\mu}\Theta)(\nabla_{\nu}\Theta)-\frac{1}{2}g_{\mu\nu}(\nabla_{\lambda}\Theta)(\nabla^{\lambda})
\end{eqnarray}
and $T^{RDE}_{\mu\nu}$ is energy-momentum tensor of  Ricci DE represented by
\begin{eqnarray}
T^{RDE}_{\mu\nu}=(\rho_{x}+p_{x})U_{\mu}U_{\nu}+p_{x}g_{\mu\nu}.
\end{eqnarray}
The term $\rho_{x}$ is the Ricci dark energy density, $p_{x}$ is the pressure of the DE and $U_{\mu}=(1,0,0,0)$ is the four velocity.
The $00$ component of Eq.(6) gives
\begin{eqnarray}
G_{00}+l C_{00}=8\pi G(T^{RDE}_{00}+T^{\Theta}_{00}).
\end{eqnarray}
For the metric, given in Eq.(2), we evaluate the following required quantities as
\begin{eqnarray}
G_{00}&=&-3(\frac{{\dot{a}^{2}}}{a^{2}}+k c^{2}),
\\
C_{00}&=&0,
\\T^{RDE}_{00}&=&\rho_{x}=\frac{3\alpha}{8\pi}(\frac{\ddot{a}}{a^{3}c^{2}}+\frac{k}{a^{2}}),
\\T^{\Theta}_{00}&=&\frac{1}{2}\dot{\Theta}^{2}.
\end{eqnarray}
From the field equation associated with the scalar field, the Pontryagin term $^{*}RR$ is zero for the given metric, thus we get
\begin{eqnarray}
g^{\mu\nu}\nabla_{\mu}\nabla_{\nu}\Theta= g^{\mu\nu}[\partial_{\mu}\partial_{\nu}\Theta-\Gamma^{\lambda}~_{\mu\nu}\partial_{\lambda}\Theta]=0.
\end{eqnarray}
After evaluating the Christoffel symbols for the metric (2), we use these values in last equation and have the following second order differential equation

\begin{eqnarray}
\ddot{\Theta}+4\frac{\dot{a}}{a}\dot{\Theta}=0,
\end{eqnarray}
which yields the solution, given as
\begin{eqnarray}
\dot{\Theta}=K a^{-4},
\end{eqnarray}
where $K$ is constant of integration. When we make use of all values in Eq.(15), we arrive at
\begin{eqnarray}
\frac{\dot{a}^{2}}{a^{2}}+k c^{2}=-[\alpha G(\frac{\ddot{a}}{a^{3}c^{2}}+\frac{k}{a^{2}})+\frac{4}{3}\pi G K^{2}a^{-8}].
\end{eqnarray}
Using the gravitational units $c=G=1$, the last equation takes the form
\begin{eqnarray}
\frac{\dot{a}^{2}}{a^{2}}+k=-[\alpha (\frac{\ddot{a}}{a^{3}}+\frac{k}{a^{2}})+\frac{4}{3}\pi K^{2}a^{-8}].
\end{eqnarray}
Now, we consider the flat AFRW universe, i.e., $k=0$, the last equation becomes
\begin{eqnarray}
\frac{\dot{a}^{2}}{a^{2}}=-[\alpha\frac{\ddot{a}}{a^{3}}+\frac{4}{3}\pi K^{2}a^{-8}].
\end{eqnarray}
Putting $\beta=\frac{4}{3}\pi K^{2}$, we have
\begin{eqnarray}
\alpha\frac{\ddot{a}}{a^{3}}+\frac{\dot{a}^{2}}{a^{2}}+\beta a^{-8}=0.
\end{eqnarray}
To find scale factor $a(t)$, we make the following substitutions
\begin{eqnarray}
u(a)=\frac{da}{dt}~~~~\Rightarrow~~~~u\frac{du}{da}=\frac{d^{2}a}{dt^{2}}.
\end{eqnarray}
Using Eq.(27) in Eq.(26), we obtain
\begin{eqnarray}
u\frac{du}{da}+\frac{a}{\alpha}u^{2}+\frac{\beta}{\alpha}a^{-5}=0.
\end{eqnarray}
Again the substitution
\begin{eqnarray}
u^{2}(a)=V(a)~~~~~\Rightarrow~~~~~u(a)\frac{du}{da}=\frac{1}{2}\frac{dV}{da}
\end{eqnarray}
makes Eq.(28) more tranquil, given as
\begin{eqnarray}
\frac{dV}{da}+\frac{2a}{\alpha}V=-\frac{2\beta}{\alpha}a^{-5}.
\end{eqnarray}
It is a linear ordinary differential equation and its straightforward solution can be written, in term of $V(a)$, as
\begin{eqnarray}
V(a)=\frac{\beta}{2\alpha}\frac{1}{a^{4}}.
\end{eqnarray}
On backward substitution in Eq.(29) and then in Eq.(27), we obtain
\begin{eqnarray}
u(a)=\frac{da}{dt}=\sqrt{\frac{\beta}{2\alpha}}\frac{1}{a^{2}}.
\end{eqnarray}
Solving this differential equation by using separation of variables method, the scale factor $a(t)$ is explored as
\begin{eqnarray}
a(t)=[{\sqrt{\frac{\beta}{2\alpha}}3t+C_{0}}]^{\frac{1}{3}},
\end{eqnarray}
where $C_{0}$ is constant of integration. Putting the value of $\beta=\frac{4}{3}\pi K^{2}$  and using $\alpha\simeq\frac{1}{2}$
as found in \cite{[16]}, the last equation takes the form
\begin{eqnarray}
a(t)=[2\sqrt{3\pi}K t+C_{0}]^{\frac{1}{3}}.
\end{eqnarray}
Using this value of scale factor in Eq.(4), we obtain the Ricci DE as
\begin{eqnarray}
\rho_x=-\frac{K^2}{2}[2\sqrt{3\pi}K t+C_{0}]^{-\frac{8}{3}}.
\end{eqnarray}

\section{Discussion}
This paper is devoted to explore the Ricci DE of AFRW universe in the framework of CS modified gravity
by finding the scale factor. It turned out to be well-defined and definite. Further, the graphical comparison
of the Ricci DEs as well as the scale factors among GCG, FRW and AFRW models are presented.

The scale factors of FRW \cite{[17]} and the GCG \cite{[18]} models are respectively, given as
\begin{eqnarray}
a(t)&=&(\frac{2\beta}{3 c_{1}})^{\frac{1}{6}}\sinh^{\frac{1}{3}}(3\sqrt{c_{1}}t),\\
a(t)&=& a_{0}\sinh^{\frac{2}{3}}\sqrt{6\pi A}t,
\end{eqnarray}
where $\beta=\frac{4\pi C^{2}}{3}.$ We allocate the following values to the different parameters involving in Eq.(34), Eq.(36) and Eq.(37)
\begin{eqnarray} K=\frac{1}{70},~~C_{0}=\frac{1}{500},C=\frac{\sqrt{3}}{100},~c_{1}=\frac{1}{500},a_{0}=1,~~A=(\frac{1}{35})^{2}.
\end{eqnarray}
Now, we plot a graph of the scale factors given in Eq.(34), Eq.(36) and Eq.(37).
The combined graph, for three different cases, is distinguished  as
FRW (yellow line), AFRW (black line) and GCG (Red line) in figure (1). The graphical comparison shows that
our result match exactly with that of FRW for particular choice
of the constant of integration, involving in their respective scale factors.
\begin{figure}[H]
\centering
\includegraphics[width=3in]{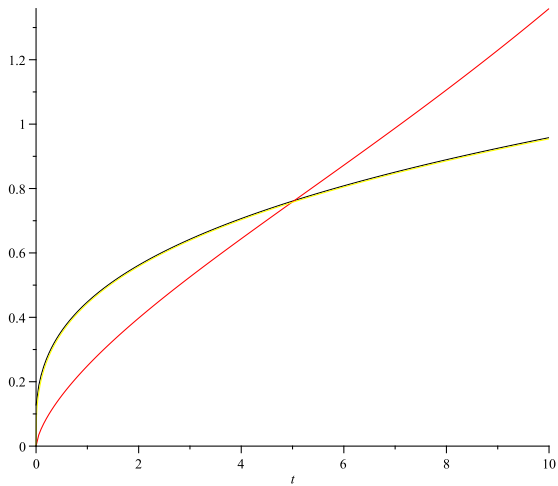}\\
\caption{\small{Scale factors of AFRW, FRW and GCG}}\label{pic01}.
\end{figure}
The authors \cite{[17]} argued that the results obtained for flat FRW
model, in CS gravity, are similar to those of GCG, available in literature \cite{[18]}. Myung \cite{[22]} commented that the
Ricci DE in CS modified gravity reduces to Ricci DE with a minimally coupled scalar, The similarity
between FRW universe and GCG is limited to the de Sitter phase derived by the cosmological
constant in the future. Thus, the same conclusion will be true for the case of AFRW model, as the results of AFRW are similar to those of FRW model.

The Ricci DE density of FRW \cite{[17]} and the energy density of GCG \cite{[18]} models are given as
\begin{eqnarray}
\rho_x&=&-c_2(\frac{1}{9}\coth^2{3\sqrt{c_1}t}+\frac{1}{3}),
\\
\rho(t)&=& A \coth^{2}(\sqrt{6 \pi GA}t).
\end{eqnarray}
For the following particular values of the parameters involving in Eqs.(35), (39) and (40)
\begin{eqnarray}
A= \frac{1}{\sqrt{2}},~~~ G=1,~~~ K=\frac{1}{12},~~~ C_0= \frac{1}{500},~~~ c_1=\frac{1}{\sqrt{7}},~~~c_2=\frac{1}{9}
\end{eqnarray}
the graph among three densities is plotted as
\begin{figure}[H]
\centering
\includegraphics[width=3in]{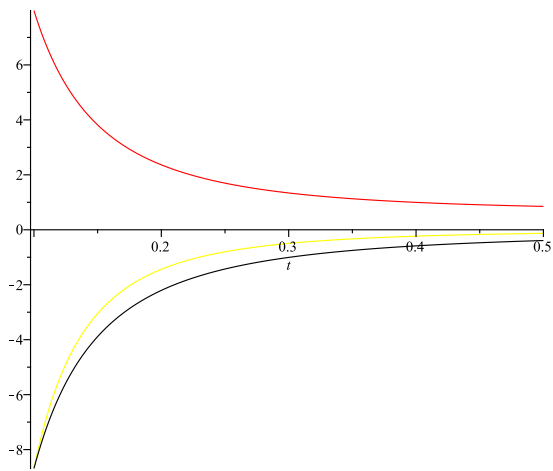}\\
\caption{Density graph of AFRW (yellow line), FRW (black line) and GCG (Red line)}\label{pic02}.
\end{figure}

This graph shows that the Ricci DE density for both FRW and AFRW models, in CS gravity, increases with time while the energy density
of GCG decreases.

\vspace{0.5cm}

{\bf Acknowledgment}
We acknowledge the remarkable assistance of the Higher Education
Commission Islamabad, Pakistan, and thankful for its financial
support through the {\it Indigenous PhD 5000 Fellowship Program
Batch-III}.

\end{document}